\documentclass[12pt]{revtex4}
\usepackage{amssymb}
\usepackage{amsmath}
\newcommand{\be}{\begin{equation}}
\newcommand{\ee}{\end{equation}}
\DeclareMathOperator{\Tr}{Tr}

\begin{document}

\title{Nonlinear unitary quantum collapse model with self-generated noise}

\author{Tam\'as Geszti}
\email{geszti@elte.hu}
\affiliation{Department of Physics of Complex Systems, E\"otv\"os Lor\'and University, 
Budapest, Hungary}
\begin{abstract}
Collapse models including some external noise of unknown origin are routinely used to describe phenomena on the quantum-classical border; in particular, quantum measurement. Although containing nonlinear dynamics and thereby exposed to the possibility of superluminal signaling in individual events, such models are widely accepted on the basis of fully reproducing the non-signaling statistical predictions of quantum mechanics. Here we present a deterministic nonlinear model without any external noise, in which randomness - instead of being universally present - emerges in the measurement process, from deterministic irregular dynamics of the detectors. The treatment is based on a minimally nonlinear von Neumann equation for a Stern-Gerlach or Bell-type measuring setup, containing coordinate and momentum operators in a self-adjoint skew-symmetric, split scalar product structure over the configuration space. The microscopic states of the detectors act as a nonlocal set of hidden parameters, controlling individual outcomes. The model is shown to display pumping of weights between setup-defined basis states, with a single winner randomly selected and the rest collapsing to zero. Environmental decoherence has no role in the scenario. Through stochastic modelling, based on  Pearle's "gambler's ruin" scheme, outcome probabilities are shown to obey Born's rule under a no-drift or "fair-game" condition. This fully reproduces quantum statistical predictions, implying that the proposed non-linear deterministic model satisfies the non-signaling requirement. Our treatment is still vulnerable to hidden signaling in individual events, which remains to be handled by future research.

\end{abstract}

\maketitle

\section{Introduction}
Collapse phenomenology accurately describes a huge variety of quantum measurements. The paradigmatic example is Stern-Gerlach (SG)  spin measurement, in which a collimated particle beam entering in an unknown spin state is separated into non-overlapping partial beams, each carrying one of pre-defined spin components. Separation is followed by recording signals from a set of detectors, each of them covering one of the partial beams. For each incoming particle, apart from noise, the observed signals are in total anticorrelation: one detector, wich is selected randomly, fires, the others not. Probabilities are controlled by the incoming amplitudes according to Born's rule; the partial beam corresponding to the  "winning" spin component becomes the initial state for further evolution, with all the rest disappearing; that is meant by the widespread term "quantum collapse" \cite{whz}. 

Collapse is not portrayed by the solutions of a linear Schr\"odinger equation; mainly because linearity implies that the coefficients of a superposition remain constant during evolution, hence by Born's rule, the statistical weight of each of the separated branches is conserved, leaving no way for one growing, others disappearing. The same conclusion is reached by von Neumann \cite{vNbook} in the language of the density matrix: linear dynamics of the density matrix keeps the statistical weights characterizing the thermal state of the apparatus unchanged for indefinite time, leaving no way for Born's rule probabilities to emerge. That offers the choice: to accept the distinguished status of the measurement process - the Copenhagen interpretation - or to find the proper nonlinear extension of the Schr\"odinger equation. 
Nonlinearity opens a way from deterministic dynamics to emergent quantum randomness; instead of through the large number of uncontrolled degrees of freedom as thought earlier, rather through the sensitivity of nonlinear dynamics to initial conditions, as convincingly illustrated by low-dimensional models of classical chaos \cite{chaos}. 

Statistics are supported by a selection of initial states. Bell inequality measurements \cite{bell,brunner}, as exposed in a particularly clear way by the CHSH argument \cite{chsh}, demonstrate that those initial states cannot be carried by the incoming particles, be it any kind of local hidden parameters (in this context "local" means that the particles are from a common source) \cite{hensenhanson}. That directs one towards the alternative to seek the relevant hidden parameters in the enormous pool of random initial states offered by the macroscopic measuring apparatus containing a number of distant detectors, described by a nonlocal set of multidimensional hidden parameters, thereby making detectors part of "reality". That approach violates the optimistic expectation to keep the process of observation outside the physics one wants to observe; that expectation is satisfied throughout classical physics, however, violated by the Copenhagen interpretation anyway. The alternative followed here is to lift the abstract status of the measurement process, making it part of the physics observed, displaying Born's rule as a consequence of deterministic dynamics, instead of being an independent part of the law.

The relation of all that to special relativity is nontrivial. Within the Copenhagen interpretation, a quantum state determines the statistics of measurement outcomes; it has no relation to individual events. That restriction grants "peaceful coexistence" of linear Schr\"odinger-Born quantum mechanics and relativity \cite{shimony}. Involving nonlinearity may strain the situation \cite{gisinhelv,signaling}: nonlinear dynamics carries the danger of superluminal signaling between remote partners \cite{nogispol}. Again, making distinction between individual events and measurement statistics is an important issue in clarifying the situation.

Including nonlinear terms in the Schr\"odinger equation had long been an obvious scope of research \cite{bbmw,vecsi,ns}. Collapse models based on some postulated external noise of unspecified origin, most of them using white noise for efficient modeling under the name Continuous Spontaneous Localization (in what follows: CSL) \cite{csl}, soon became a  dominant direction of research. Such models can fully reproduce Born's rule statistics of measurement outcomes, which is sufficient to get around the catch of superluminal signaling \cite{brunner}, as long as individual events are not considered. Below we show that deterministic noisy behavior emerging from nonlinear dynamics of particle detectors is capable of the same performance, without resorting to any external source of noise \cite{bohmian}. 

As presented below in detail, the basic difference between our treatment and CSL - possibly, the price for equal performance - is that whereas those models postulate an omnipresent external noise, we restrict randomness in the measurement process to chaotic dynamics of the detectors. In the analysis to follow, that is expressed in the particular choice of the basis on which collapse reveals itself: it is defined by the different possible outcomes of a single measurement event \cite{tocsl}.

Unfortunately, our approach, just like those noise-based collapse models, by no means excludes signaling present in individual events, excluding to consider it a flawless account for the dynamics of quantum measurement. However, "for all practical purposes", one can go back to the Copenhagen spirit, accepting to consider individual events insignificant from the point of view of Born's rule based protocol. What has still been gained is to get away without postulating an unknown external field.

To derive Born's rule is a nontrivial part of the task. Within the framework of noise-based, CSL-type collapse models, random selection is a dynamical process, efficiently modeled by a zero-sum, fair game of the "gambler's ruin" type \cite{gambler}, in which the gamblers are measurement-defined basis states, entering with their statistical weights as stakes, the sum of which is kept constant ("zero-sum game"). If a gambler looses its stake, it drops out of the game. Collapse - one gambler winning all, the rest loosing all - results from a sequence of small steps of unbiased coin tossing - probabilities 1/2, 1/2 each time for a randomly chosen couple of gamblers - to make it a "fair game". With that property granted, Born's rule statistics of the final outcomes is traced back to incoming amplitudes controlling the {\sl initial} state of the otherwise neutral selection process. Below we use the same approach, starting from a deterministic model. Stochastic modelling on a later stage of the analysis helps getting some more insight to the dynamics of the collapse process \cite{Pearle}.

Our treatment is focused on typical measurement situations; analysing non-measurement scenarios on the quantum-classical border, amply discussed in terms of collapse models, may require a different approach.

\section{The framework}

In what follows, we are looking for strictly deterministic evolution of an individual system, composed of a microsystem and a delocalized set of detectors labelled by $d$,  the whole being characterized by density matrix $\hat{\boldsymbol\varrho}$ evolving according to a nonlinear Von Neumann equation
\be\begin{split}\label{vNnonlin}
 \partial_t \hat{\boldsymbol\varrho} &= 
                 -\frac{i}{\hbar}\left[\hat H(\hat{\boldsymbol\varrho}),\hat{\boldsymbol\varrho}\right] \\
         =&  -\frac{i}{\hbar} \left[ \left(\hat H_0  + i\zeta 
  \left( \hat x \hat{\boldsymbol\varrho}\hat p -  \hat p \hat{\boldsymbol\varrho} \hat x \right)\right) ,\hat{\boldsymbol\varrho}\right]
\end{split}\ee
where $\hat H_0$ describes linear evolution of the incoming particles and all detectors interacting with them, including  environments of the whole measurement setup \cite{petruccione}. A minimal nonlinearity is introduced by adding a linearly $\hat{\boldsymbol\varrho}$-dependent term to the Hamiltonian, constructed to be self-adjoint, in order to preserve the norm of the density matrix during evolution. In the new term the density matrix is enclosed by a skew-symmetric combination of two canonically conjugate vector observables $\hat x $ and $\hat p $ acting on the configuration space including incoming particles, detectors and their environments. The constant $\zeta$ characterizes the strength of nonlinearity. Since the combination $\hat x \hat{\boldsymbol\varrho}\hat p - \hat p \hat{\boldsymbol\varrho} \hat x$ has the dimension of action, measured in quantum units by the factor $\hbar^{-1}$, $\zeta$ acquires dimensionality $t^{-1}$. 

The immediate insight gained therefrom is to connect the quantum-classical border to time scales. Indeed, nonlinearity effects, marking the quantum-classical border, get manifest after some characteristic time, under control of the action implied; leaving long time for microsystems, short time for macroscopic ones to live in the quantum way. As shown below, in our scheme - somewhat contrary to widespread expectations - decoherence effects \cite{petruccione,dekoh} have no direct influence on the collapse process; genuine quantum phenomena, described by the linear Schr\"odinger equation to an excellent approximation, are those running to the end before nonlinearity takes the control. In the process of quantum measurement, particle-detector interaction happens on the $\hat H_0$ timescale; then collapse follows under the control of $\mathcal{O}(\zeta)$ terms, as the action involved in detector functioning grows.

Turning back to Eq. (\ref{vNnonlin}), operator products  $ \hat x \hat{\boldsymbol\varrho} \hat p $ and  $ \hat p \hat{\boldsymbol\varrho} \hat x $ are meant to be scalar products of $\hat x$ and $\hat p$ over the full configuration space, to result in a scalar Hamiltonian; that structure obtains importance in tackling the locality issue, see Sec. 5 below. A scalar self-adjoint Hamiltonian implies unitarity; within the "gambler's ruin" collapse phenomenology \cite{gambler}, that grants that the emerging random selection of the firing detector(s) will be a zero-sum game. To keep it a fair game too, it relies on non-trivial physical conditions, as demonstrated in the course of the analysis below. Momentum is unavoidably part of the scenario; a live cat is not only elsewhere, it jumps as well.

Quantum measurement is done by a set of remote detectors connected into some measurement setup, devised to count individual detection events and/or given combinations of coincidences. Measurement typically starts by a separation process -- sometimes called the 'von Neumann measurement' -- in which the microsystem enters in the form of a collimated particle beam or a few entangled beams from a common source; then unfolded into non-overlapping, individually collimated partial waves, each reaching one of the detectors. In a steady-state scattering picture, separation results in a superposition of the form $\sum_k c_k|k\rangle$, as expanded into some steady orthonormal basis states $|k\rangle$, defined by the actual measurement setup: $|\pm\rangle$ for SG, $|\pm,\pm\rangle$ for CHSH, $|l,r\rangle$ and $|r,l\rangle$ (for "left" and "right") for the original EPR setup \cite{EPR}. Interaction with the detectors at time $t=0$ creates an entangled state of particles and detectors, in the form
\be\label{Psi}
{\bf|\Psi\rangle}~=~\sum_k c_k|k\rangle|\Phi_k\rangle,
\ee
which becomes the initial state for subsequent temporal changes -- in particular, collapse.  $|\Phi_k(t)\rangle$ denotes a multi-detector pure state vector, with local environments of each detector included. To fix ambiguity of the products in Eq. (\ref{Psi}), the multi-detector state vectors are defined to remain normalized during evolution: $\langle\Phi_k(t)|\Phi_k(t)\rangle~=~1$.

The density matrix corresponding to state vector (\ref{Psi}) takes the form 
\be\label{dnstmtrx}
\hat{\boldsymbol\varrho}~=~{\bf|\Psi\rangle\langle\Psi|}~=~
       \sum_{k,l} ~ |k\rangle\langle l|~c_kc_l^*~\hat R_{kl}(t),
\ee
where we have introduced the shorthand notation
\be\label{gammaRdef}
\hat R_{kl}~=~|\Phi_k\rangle\langle\Phi_l|.
\ee
In coordinate representation, the operator $\hat R_{kl}$ appears as a Hermitian matrix $R_{kl}( x, x';t)=R_{lk}( x', x;t)^*$, acting on the configuration space of all detectors, including their local environments \cite{positivity}. 

As already mentioned, $c_k(t)$ can change in time by nonlinear dynamics only. We are looking for a solution of Eq. (\ref{vNnonlin}) preserving the form (\ref{dnstmtrx}), with coefficients $c_k(t)$ evolving during the measurement process. Their initial values are fixed down to the incoming  $c_k$ in the $\mathcal{O}(\hat H_0)$ fast particle-detector interaction phase. It is the time dependence of the coefficients $c_k(t)$ through which statistical weights
\be\label{weight}
 w_k(t)~=~|c_k(t)|^2
\ee
are pumped by nonlinearity from one basis state to another, starting from initial values $w_k(0)=|c_k(0)|^2$; finally resulting in collapse. On the other hand, the time dependence of a state vector $|\Phi_k(t)\rangle$, while keeping detector $d$  activated or quiet according to whether the partial waves forming basis state $|k\rangle$ reach the actual $d$ or not, still depends on all microscopic details of what happens at that detector and its environment. Those details are controlled by local microscopic initial conditions drawn randomly from an enormous Gibbsian ensemble; serving as multi-local hidden parameters. Pumping and local dynamics are coupled through Eq. (\ref{vNnonlin}); we now turn to explore the way that can give rise to collapse.

\section{General structure of the solution}

In what follows, we use the interaction picture \cite{diosiprivate}, and look for a solution of Eq. (\ref{vNnonlin}) in the form of Eq. (\ref{dnstmtrx}). Normalization of the detector state vectors involves
\be\label{Rnorm}
\Tr_x\hat R_{kk}(t)~=1;~~~~\frac{d}{dt}\Tr_x\hat R_{kk}(t)~=0.
\ee
Eq. (\ref{dnstmtrx}) leaves open the way phases are shared between factors $c_k(t)c_l^*(t)$ and $\hat R_{kl}(t)$. 
Using Eq. (\ref{weight}), we take the parametrization
\be\label{phase}
c_k(t)c_l^*(t)~=~\sqrt{w_k(t) w_l(t)}~ e^{i\varphi_{kl}(t)},
\ee
and make use of the freedom to postulate
\be\label{phases}
\dot\varphi_{kl}~=~0,~~~\forall~ k,l
\ee
assigning phase changes to the operator factors $\hat R_{kl}(t)$. We note that in Eq. (\ref{Rnorm}), $\Tr_x$ denotes trace over the configuration space, not over $k$; accordingly, the mean value of any configuration-space operator $\hat A_x$ acting on the  configuration space coordinates, not on indices $k$, is
\be\label{mean}
\langle{\bf\Psi}| \hat A_x |{\bf\Psi}\rangle~=~\sum_k |c_k|^2 \langle\Phi_k|\hat A_x|\Phi_k\rangle~=~
\sum_k w_k \langle \hat A_x \rangle_k.
\ee
After that preparation, considering that the vector operators $\hat x$ and $\hat p$ are configuration-space operators as defined above, in straightforward steps one arrives at the system of equations
\be\begin{split}\label{nondiag} 
\frac{d}{dt}&\hat R_{kl}+\frac12\left(\frac{\dot w_k}{w_k}+\frac{\dot w_l}{w_l}\right) \hat R_{kl}\\
&=\frac{\zeta}{\hbar}\sum_m w_m 
\left(\langle\hat p\rangle_m \cdot \{\hat x,\hat R_{kl}\} 
- \langle\hat x\rangle_m\cdot\{\hat p,\hat R_{kl}\}\right)
\end{split}\ee
where $\{...,...\}$ denotes the anticommutator. For $l=k$, taking the configuration-space trace and using properties (\ref{Rnorm}) and (\ref{mean}) as well as cyclic invariance of the trace, one finally obtains the pumping rates in the form
\be\label{pump}
\dot w_k~=~\zeta~ \sum_m w_k w_m A_{km},
\ee
where 
\be\label{toandfro}
A_{km}(t)~=~\frac{2}{\hbar}~  \Big(\langle\hat x\rangle_k \cdot \langle\hat p\rangle_m
             ~-~ \langle\hat x\rangle_m \cdot \langle\hat p\rangle_k\Big).
\ee
Eqs. (\ref{pump}) with (\ref{toandfro}) constitute a system of balance equations, with  the obvious antisymmetry property
\be\label{antisymm}
A_{km}=-A_{mk},
\ee 
granting $\sum_k\dot w_k=0$, i.e., conservation of the total weight, obviously traced back to self-adjointness of the Hamiltonian. $A_{km}(t)$
is a fluctuating dimensionless quantity, changing in time as a consequence of chaotic microscopic motions over the whole configuration space of all detectors and their environments, according to the actual functional form of the diagonal elements $R_{kk}( x, x';t)$ which depend on all microscopic details of the state of all detectors, both those controlled by their preparation and those which are not accessible to preparation protocols. The latter constitute the set of  "multilocal hidden parameters", guiding an individual detection event towards a definite, although statistically random, individually unpredictable outcome. In the next section we model $A_{km}(t)$ by random noise to get more insight into the way that happens.

It is worth mentioning that Eqs. (\ref{nondiag}) to (\ref{toandfro}) display a clear separation of time scales based on the smallness of $\zeta$: evolution of  $\hat R_{kl}(t)$, eventually including decoherence effects, is dominated by $\mathcal{O}(\zeta^0)$ dynamics accounted for by the interaction picture -- nonlinearity adds but small corrections  to that -- whereas weights $w_k(t)$ are changing in slower, $\mathcal{O}(\zeta)$ pumping processes.

\section{Stochastic modelling}

Equation (\ref{pump}), with $A_{km}(t)$ depending on microscopic processes in a number of detectors, describes a chaotic walk of vector $\{w_k\}$ over the simplex $\sum_k w_k=1$. The factors $w_k w_m$ impose a peculiar scenario on the walk: whenever a particular component $k$ reaches zero, that component never revives any more; that continues until a single component $i$ remains at $w_i=1;~w_k=0,~\forall k\not=i$; then the walk stops there. All such points are {\sl attractors} of the walk; the particular attractor reached marks the outcome of the measurement. As mentioned in the Introduction, that scenario fits into the 'gambler's ruin' model \cite{gambler}.

While keeping in mind that all this is deterministic dynamics, Equations (\ref{pump} - \ref{toandfro}) open the way to model collapse as a stochastic process, assigning a probability $P_i$ to each possible measurement outcome. The outcome probabilities depend on the vector of initial weights; the basic task of stochastic modelling is to determine the functional form $P_i(\{w_k(t=0)\})$ of that dependence.

Having a derivation of Born's rule in mind, we only need to require that exit probabilities would not change in the course of the walk; in game theory language, that is called the {\sl martingale} property. Adapting the elementary treatment of the 'gambler's ruin' \cite{gambler}, that requirement can be checked on each step of the game, in which some amount $\Delta$ of weight is transferred between two states $m$ and $n$. The martingale property is granted if that transfer happens at equal probability $1/2$ in both directions (a  {\sl fair game}); then during the transfer, the outcome probabilities do not change. That is expressed by a system of difference equations
\be\begin{split}\label{onestep}
 &P_i\big(\{ w_k\}\big) 
      ~=~{}\textstyle\frac{1}{2}~ P_i\big(w_m+\Delta, w_n-\Delta,\{w_{k\not=m,n}\} \big) 
  \\ &  ~~~~~~~~+{}\textstyle\frac{1}{2}~ P_i\big(w_m - \Delta,w_n+\Delta,\{w_{k\not=m,n}\} \big); ~~\forall i,m,n.
\end{split}\ee
As boundary conditions to the above equations, we observe that starting right at a particular output port $j$, no choice remains but to take the same output:
\be\label{boundcond}
{\textsl for}~ w_k=\delta_{kj},~~~~~~~ P_i\big(\{ w_k\}\big)~=~\delta_{ij},~~\forall j
\ee
($\delta_{ij}$ is the Kronecker symbol). As checked by direct substitution, for any number of basis states playing the game, the solution of Eqs. (\ref{onestep}) with boundary conditions (\ref{boundcond}) is the linear one,
\be\label{Born}
P_i~=w_i,
\ee
which is Born's rule. A somewhat more sophisticated way to arrive at the same conclusion is to take the continuum limit $\Delta\to0$; then expanding Eq. (\ref{onestep}) in powers of $\Delta$, $\mathcal{O}(\Delta^0)$ and $\mathcal{O}(\Delta)$ terms cancel; dominant $\mathcal{O}(\Delta^2)$ terms give the system of backward Fokker-Planck equations \cite{gardiner}
\be\label{bFP}\left(\frac{\partial}{\partial w_m}-\frac{\partial}{\partial w_n}\right)^2 
                         P_i\big(\{ w_k\}\big)~=~0; ~~\forall~ i,m,n
\ee
which under boundary conditions (\ref{boundcond}) furnish again the Born's rule solution (\ref{Born}). It is important to mention that the above derivation is totally insensitive to all microscopic details of the detection system; the only requirement is the $\pm\Delta$ symmetry of pumping steps, entailing driftless dynamics, i.e. "fair-play" martingale gaming.

To have some additional insight to the dynamics of the walk process resulting in Born's rule statistics, we turn to more flexible models of stochastic evolution. The obvious option is to replace $A_{km}(t)$ by a random noise, stationary during the collapse process. Denoting averages over time by overline, we postulate
\be\label{noise}
\overline {A_{km}}~=~0
\ee
expressing the no-drift (martingale) property, necessary to arrive at Born's rule. Further -- although it may be far from reality due to the rapidness of quantum collapse -- in order to make the model a tractable one, we approximate $A_{km}(t)$ by white noise, quasi-stationary in time, its autocorrelation function being of the form
\be\label{autocorr}
 \langle A_{km}(0) A_{km}(t) \rangle~=~\overline {A_{km}^2}~\delta(t/\tau_{km})
\ee
with an effective correlation time $\tau_{km}$. That turns the evolution of $\{w_k(t)\}$ into a Markovian diffusion process; using the Green-Kubo relation \cite{greenkubo} $D_{km}=\int_0^\infty  \langle \dot w_k(0) \dot w_m(t) \rangle dt$, one obtains the diffusion coefficients along each axis $\{km\}$ in the form 
\be\label{diffcoeff}
D_{km}~=~d_{km}(w_k w_m)^2 ; ~~d_{km}~=~ 4\zeta^2  \tau_{km} \overline {A_{km}^2}.
\ee
Finally, taking into account that according to Eq. (\ref{noise}), noise brings no drift into the process, we obtain the time-dependent ("forward") Fokker-Planck equation
\be\begin{split}\label{FP}
\frac{\partial}{\partial t} &P\big(\{ w_k\},t\big)~=~\sum_{mn}d_{mn} \\&
         \left(\frac{\partial}{\partial w_m}-\frac{\partial}{\partial w_n}\right)^2 
                      \Big[(w_k w_m)^2 P\big(\{ w_k\},t\big)\Big].
\end{split}\ee
where the particular combination of derivatives on the r.h.s., also appearing in Eq. (\ref{bFP}) \cite{finitetime}, is traced back to the antisymmetry property (\ref{antisymm}), to be preserved when turning to stochastic modeling. 

The factors $(w_k w_m)^2$ in Eq. (\ref{FP}) specify the character of output ports $w_i=1,~w_{k\not=i}=0$: those are not absorbing, rather "natural boundaries" in the sense that as diffusive motion of the vector $\{ w_k\}$ is approaching one of those points, dynamics slows down to finally stop there \cite{natbound}; i.e., the system is collapsing into the corresponding basis state. To see in some detail how that happens, we focus on the two-port (Stern-Gerlach) case. Defining $w_1=w,~ w_2=1-w,~d_{12}=\mu/4$, the equation for $P(w,t)$ reads
\be\label{SGequation}
\dot P~=~\mu~[w(1-w)]^2~P'',
\ee
with the analytical solution for diffusion starting from initial value $w_0=|c_1(t=0)|^2$ \cite{Pearle} 
\be\label{solution} 
P(w,t)~=~\sqrt{\frac{w_0(1-w_0)}{w^3(1-w)^3}} \frac{e^{-\frac{\mu t}{4}}} {\sqrt{4\pi\mu t}}
             e^{-\frac{1}{4\mu t}[\ln\frac{w}{1-w}-\ln\frac{w_0}{1-w_0}]}
\ee
that starting from initial condition $\delta(w-w_0)$ for $t\to 0$, subsequently splits into two peaks, first broadening, then shrinking on the two output points $w=0,1$ with the respective weights $w_0$ and $1-w_0$.

\section{The locality issue}

To proceed further, we assume that before measurement starts, remote detectors are uncorrelated. That imposes that following interaction with the separated partial waves of incoming particles from a common source, the multi-detector density matrices $\hat R_{kl}$ in Eq. (\ref{dnstmtrx}) take the direct product structure 
\be\label{dirprod}
\hat R_{kl}(t) ~=~ \bigotimes_d \hat r_{kl}^d(t),
\ee
the one-detector density matrix elements being built of normalized one-detector states as 
$\hat r_{kl}^d=|\varphi_k^d\rangle\langle\varphi_l^d|$. That structure is preserved during subsequent evolution, since the Hamiltonian governing the evolution according to Eq. (\ref{vNnonlin}) is decomposed into a sum over detectors:
\be\label{sumoverdet}
\hat H(\hat{\boldsymbol\varrho}) = \sum_d  \left(\hat H_0^d  + i\zeta \left( \hat x_d \hat{\boldsymbol\varrho} \hat p_d -  \hat p_d \hat{\boldsymbol\varrho} \hat x_d \right)\right).
\ee
Here $\hat H_0^d$ describes linear evolution of detector $d$, initially interacting with the respective partial wave of incoming particles, subsequently with its environment. The separation of the $\hat{\boldsymbol\varrho}$-dependent term into a sum over detectors $d$ needs explanation. As mentioned above, the operator products  $ \hat x \hat{\boldsymbol\varrho} \hat p $ and  $ \hat p \hat{\boldsymbol\varrho} \hat x $ in Eq. (\ref{vNnonlin}) are meant to be scalar products of $\hat x$ and $\hat p$ over the full configuration space to result in a scalar Hamiltonian. It is that scalar product structure which entails decomposition into a sum over $d$, as displayed in Eq. (\ref{sumoverdet}); each term is a direct product of an operator acting on one subsystem, and unit operators on all the others \cite{peres}. Substitution of Eqs. (\ref{dirprod}) and (\ref{sumoverdet}) into Eq. (\ref{vNnonlin}), following steps preceding Eq. (\ref{nondiag}), including use of the interaction picture, results in a sum over $d$ of the analogous one-detector equations
\be\begin{split}\label{onedet} 
\frac{d}{dt}&\hat r_{kl}^d=-\frac12\left(\frac{\dot w_k}{w_k}+\frac{\dot w_l}{w_l}\right) \hat r_{kl}^d   
 \\&+\frac{\zeta}{\hbar}\sum_m w_m 
\left(\langle\hat p_d\rangle_m \cdot \{\hat x_d,\hat r_{kl}^d\} 
- \langle\hat x_d\rangle_m\cdot\{\hat p_d,\hat r_{kl}^d\}\right),
\end{split}\ee
for each $d$ multiplied by a factor $\bigotimes_{d'\not=d}r_{kl}^{d'}$. The pumping rate equation (\ref{pump}) remains unchanged, however, the noise terms appear as a sum over detectors,
\be\label{noisesum}
A_{km}(t)~=~\sum_d A_{km}^d~=~\frac{2}{\hbar}~\sum_d  \Big(\langle\hat x_d\rangle_k \cdot \langle\hat p_d\rangle_m
             ~-~ \langle\hat x_d\rangle_m \cdot \langle\hat p_d\rangle_k\Big)
\ee
Accordingly, pumping is added up from contributions of the individual detectors; as seen from Eq. (\ref{noisesum}), summing contributions to Eq. (\ref{pump}) over $k$ gives zero for each detector $d$ separatly, expressing a kind of "detailed unitarity". 

It is important to note that whereas Eqs. (\ref{onedet}) are quasi-local, since each of them refers to a single detector $d$, they include the global weights $w_m$, representing remote entanglement in the scenario.  It remains to be shown that those global weights do not allow signaling through statistics. As mentioned in the Introduction, that task is solved through the results summarized in Ref. \cite{brunner}: since our treatment reproduces Born's-rule-based quantum statistical predictions, it remains a subset of non-signaling theories. Thereby, the performance of our model is equivalent to that of CSL-type models \cite{csl}, with the advantage that here noise appears as a consequence of multilocal dynamics of the measurement apparatus.

What remains to be explained is the nonlocality of individual events, not accessible through measurement statistics but still part of reality. Although having demonstrated non-signaling as far as statistics is concerned, "spooky action-at-a-distance" remains in individual events. The same is true about CSL: both approaches furnish but extensions of Shimony's "peaceful coexistence between quantum mechanics and relativity" \cite{shimony}. Therefore statistics - beyond being the only predictable  characteristics of what is observed - still remains unavoidably an interpretation in the philosophical sense. To get rid of that, and find a description covering individual events as well and still free of nonlocality puzzles, may remain a target for future research; locating the right kind of nonlinear wave equation, local in configuration space, seems to be a promising target. The present model, like CSL-type ones, can prove to be an approximate property of the so far unknown law, not the law itself; just sufficient to pave the way to Born's rule.
 
\section{Discussion}

Taking the above analysis as a new kind of collapse model, it is an obvious scope to estimate constant $\zeta$ characterizing the strength of nonlinearity. Photon detectors, just crossing the quantum-classical border, seem to be the right object to focus on. A typical avalanche photon detector \cite{photondet} mobilizes charge $Q$ of some $2\cdot10^7$ electrons under a bias  $U=3000~ V$ during $t=10^{-8} s$, involving the dimensionless action $Q\cdot U\cdot t/\hbar \approx 10^{18}$. Within the same time $t$, that reaches classicality - as required from a detector - if $~\zeta \approx \hbar/(Q~U~t^2) = 10^{-10} ~s^{-1}$; that is our first estimate for quantum nonlinearity \cite{plancktime}. 

Eq. (\ref{noisesum}) contains products of averages evaluated in different states of a detector; coherences do not appear here, demonstrating that according to the present model, decoherence \cite{petruccione} has no role in collapse, contrary to common belief.  Instead, one can obtain some useful insight by considering detector operation as metastable dynamics \cite{metastable}. Fully developed firing is preceded by a {\sl pre-firing period}, during which the  detector, already activated by the incoming particle, in its quasi-random walk on a complicated potential surface has not yet found the passage for the avalanche towards unlimited growth to macroscopic observability. During that period, motion is like in a highly excited, quasi-stationary bound state, traveling chaotic paths, bounded both in coordinate, oscillating around a c.o.m., and momentum, oscillating around zero. That is just the behavior required for noise, driving a purely diffusive process with no drift, making fair game possible. As soon as the avalanche starts, the noise level (\ref{autocorr}) begins to grow rapidly, making the diffusion coefficients time-dependent, and eventually spoiling Born's rule statistics. That may offer an explanation to the practice of detector bias threshold, through the possible requirement that collapse - as modelled by the "gambler's ruin" game - should run to conclusion each time during the metastable pre-firing period. 

To sum up, we have used the track beaten by CSL-type theories \cite{csl} to argue that a deterministic, nonlinear dynamical equation can be capable to reproduce Born's rule quantum statistics, without signaling. In several other respects, the present approach is opposite to those theories. They postulate some external source of noise; here noise is emerging from chaotic dynamics. Randomness, as quantified in Born's rule, emerging ubiquitously in CSL, appears as a characteristic of genuine measurement situations here. Our conclusions refer to all kinds of delocalized measurements on entangled states of multiparticle systems from a common source, including the Bell setup \cite{bell}; although this scenario is not "locally realistic" in the sense that local common causes do not determine the individual outcomes of quantum measurements, they do control the correlated measurement statistics of entangled remote subsystems, through local events, still under the control of global weights. Randomness is brought into the process by chaotic dynamics of the macroscopic detectors, running in remote places, with microscopic fluctuations at each of them acting as "hidden parameters" in the Bell narrative; therefore the scenario proposed here can be classified as {\sl multilocal-realistic}.

An interesting prediction of noise-based collapse models, heating due to explicit breaking of time translation symmetry by external white noise \cite{dlheat}, is absent in the present scheme: noise generation being attached to strongly chaotic dynamics of the detectors in the pre-firing period, this treatment offers no reason for heating in non-measurement situations. It is probably too early to consider negative results of attempted detection of the heating effect \cite{noheat} as a strong argument in favour of the present treatment, as contrasting CSL-type theories.

As an outlook, a few possible directions of further research should be mentioned. First, spatially separated superconducting photon detectors \cite{supercond} integrated into a coherent quantum circuit may offer a way to extend the study to correlated detectors with controllable relative phases. Second, in a non-measurement situation - in which decoherence may take a role - the noise that would accompany pumping weights between diverging modes, with its final slowing down caused by decaying global weight factors $w_k(t)$, can be detected, eventually related to the ubiquitous, still poorly understood 1/f noise \cite{oneoveref}. Finally, the treatment presented here may have a word to the elusive relationship between quantum mechanics and general relativity \cite{Kiefer}. Quantum randomness emerging from a single dynamical law: that encourages one to consider non-quantized gravity, leaving no place for intrinsic metric fluctuations.

{\bf Acknowledgment.} It is my pleasure to acknowledge that continuing discussions over many years with Lajos Di\'osi had a decisive role in getting this paper to a conclusion. I thank helpful correspondance with Nicolas Gisin, Marek Czachor and Philip Pearle, and stimulating discussions with Andr\'as  Bodor, Andr\'as Csord\'as and P\'eter Vecserny\'es.


\begin{thebibliography}{99}
\bibitem{tocsl}This distinction between CSL and our treatment has been clarified due to a recent discussion with Lajos Di\'osi, and a remark of one of my Referees.
\bibitem{positivity} Positivity of the density matrix during evolution is granted by the use of pure states.
\bibitem{finitetime} That combination appears in Ref. \cite{Pearle} with an open exponent $r$; $r>1$ is required for collapse to be concluded within a finite time. Our treatment gives $r=2$ based on dynamics, satisfying that requirement.
\bibitem{plancktime} Inverse Planck time, $10^{49}~s^{-1}$, would make nonlinearity dominant even on the subatomic scale; that seems to exclude any cosmological background.
\bibitem{metastable} The idea to characterize detectors through their metastability appears again and again over the years. A particularly clear presentation is in Ref. \cite{vankam}; see also \cite{NSgeszti}.
\end{thebibliography}
\end{document}